\documentclass{article}

\usepackage[final]{nips_2016}

\usepackage[utf8]{inputenc} 
\usepackage[T1]{fontenc}    
\usepackage{hyperref}       
\usepackage{url}            
\usepackage{booktabs}       
\usepackage{amsfonts}       
\usepackage{nicefrac}       
\usepackage{microtype}      
\usepackage{graphicx}

\newcommand{\1}{|}
\newcommand{\0}{\bigcirc}

\title{Spiking memristor logic gates are a type of time-variant perceptron}

\author{
  Ella M.~Gale\thanks{ella.gale@gmail.com} \\
  School of Experimental Psychology\\
  University of Bristol\\
  12a Priory Road, Bristol, UK BS8 1TU\\
  \texttt{ella.gale@bristol.ac.uk} \\
}

\begin{document}

\maketitle

\begin{abstract}
Memristors are low-power memory-holding resistors thought to be useful for neuromophic computing, which can compute via spike-interactions mediated through the device's short-term memory. Using interacting spikes, it is possible to build an AND gate that computes OR at the same time, similarly a full adder can be built that computes the arithmetical sum of its inputs. Here we show how these gates can be understood by modelling the memristors as a novel type of perceptron: one which is sensitive to input order. The memristor's memory can change the input weights for later inputs, and thus the memristor gates cannot be accurately described by a single perceptron, requiring either a network of time-invarient perceptrons or a complex time-varying self-reprogrammable perceptron. This work demonstrates the high functionality of memristor logic gates, and also that the addition of theasholding could enable the creation of a standard perceptron in hardware, which may have use in building neural net chips.
\end{abstract}

\section{Introduction}


Memristors are a novel electronic component discovered in 2008[1], which function as a resistor with memory. Memristors have been associated with neurons as: a description of synaptic learning [2,3], used to produce spike-time-dependent plasticity in neural network simulation[4,5], and used to model ion channels in the Hodgkin-Huxley model of neural cell membranes[6,7]. As they are low power and combine memory and processing in one unit, they have been suggested for neuromorphic computing[1]. Memristors also exhibit spiking behaviour, however, this aspect of their behaviour has been largely ignored by neuromorphic engineers, although the spiking behaviour has been used to make logic gates. In this paper, I compare these logic gates to perceptron models, and demonstrate that the spiking memristor logic requires a novel type of perceptron in order to describe their operation. 

\begin{figure}[h]
  \centering
  \fbox{\rule[-.5cm]{0cm}{4cm} 
  \includegraphics[width=0.9\linewidth]{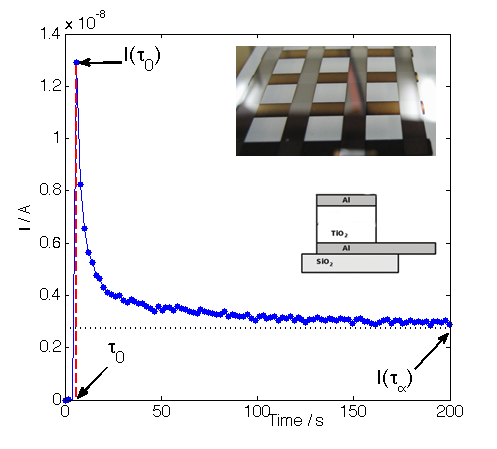}
  
  \rule[-.5cm]{0cm}{0cm}}
  \label{fig:Mem}
  \caption{Memristor `short-term memory' is encoded in its spiking current response to a voltage change.  A continuous voltage is input at time $\tau_0$ and gives a response spike in the direction of the voltage change. After 200s the current is approaching its minimum value ($I(\tau_{\infty})$ and the device is considered `zeroed', within this timeframe the memoristor has a short-term memory and a second input spike will interact with the stored value in a way that can do computation. Inset top: photograph of freshly made memristors. Inset bottom: memristor device schematic.}
\end{figure}

Much research has focussed on the long-term-memory-holding ability of the memristor, with applications in computer hard-drive and RAM memory[8]. The memristor's spiking behaviour has been compared to its short-term memory[9,10]. Biological neurons produce voltage spikes from the application of charge through a ion-gating protein (the proteins modelled as memristors), thus generate voltage spikes from current-influx; the memristors generate current spikes from a voltage application, so operate as the dual of neurons (which has led to some investigations into combining them electronically[11]). Investigations of a particular memristor, that shown in the inset to figure~\ref{fig:Mem}, and made of thin-layer of titanium dioxide which can change its resistance between aluminium electrodes[12], have yielded several interesting properties which can be used for spike-based computation. Memristors have the property that their state is a function of the integral of the charge that has gone through them--and thus non-linear. Thus, the output of a memristor is charge-dependent, and this appears as a time-variance in their operation as the charge is time-varying, and, in the case of these memristors, diminishing. Both the time-variance of the output and its non-linear diminishing nature are illustrated in figure~\ref{fig:Mem}, this property has been called `diminishing returns'. As these spikes are easy to produce, repeatable and reliable, I suggest that the spiking memristor would be a good choice for building spiking neural networks in hardware. Another aspect of their behaviour is the so-called `bounce-back' property, where it appears switching voltage polarity (or turning off the voltage) causes resulting spike--allowing outputs to be separated from inputs and read out.


Neural networks can be described as a network of perceptrons, which are loosely based on what is believed to be biological neuron operation. A perceptron performs 3 operations:
\begin{enumerate}
 \item summation, $\sum$, of inputs, $x_i$, multplied by weights $w_i$: 
  \begin{equation}
 a = \sum x_i w_i\; ;
 \label{eq:a}
 \end{equation}
 \item threasholding the output by application of a bias $\pm b$;
 \item response to the threashold, $\int$, which takes in a real-valued number, $a\pm b$, where $-\infty < a\pm b < +\infty$, which produces a spike if and only if $a \pm b$ is positive, i.e.
 \begin{eqnarray}
 \int(a\pm b) & = & \1 \;, \mathrm{if} \: a\pm b > 0\\
 \int(a\pm b)	 & = & \0 \;, \mathrm{if} \: a\pm b \leq 0 \;, 
 \end{eqnarray}
\end{enumerate}   
where $\1$ is the symbol we shall use for logical 1 (i.e. the percetron does spike) and $\0$ is the symbol for logical 0 (the perceptron does not spike). Loosely, the memristor properties above can be applied to the operation of the perceptron. The inputs are summed as a weighted sum, where the weights describe the `diminishing returns', and the output spikes can be associated with the response to the threashold. To use the memristor as a perceptron, a threshold would need to be applied. In this paper, we examine how the weights are applied and the response calculated in two single memristor gates: a spiking memristor logic gate, SpMLG[13],  which performs AND and OR operations and the spiking memristor arithmetical full adder SpMAFA[14]. Note, that these spiking logic gates are made from a single memristor, which has a single input wire, and thus the inputs values are input sequentially (i.e. separated in time rather than space).

\section{A simple single memristor spiking logic gate: AND}

\begin{figure}[ht!]
  \centering
  \fbox{\rule[-.5cm]{0cm}{4cm} 
  \includegraphics[width=0.9\linewidth]{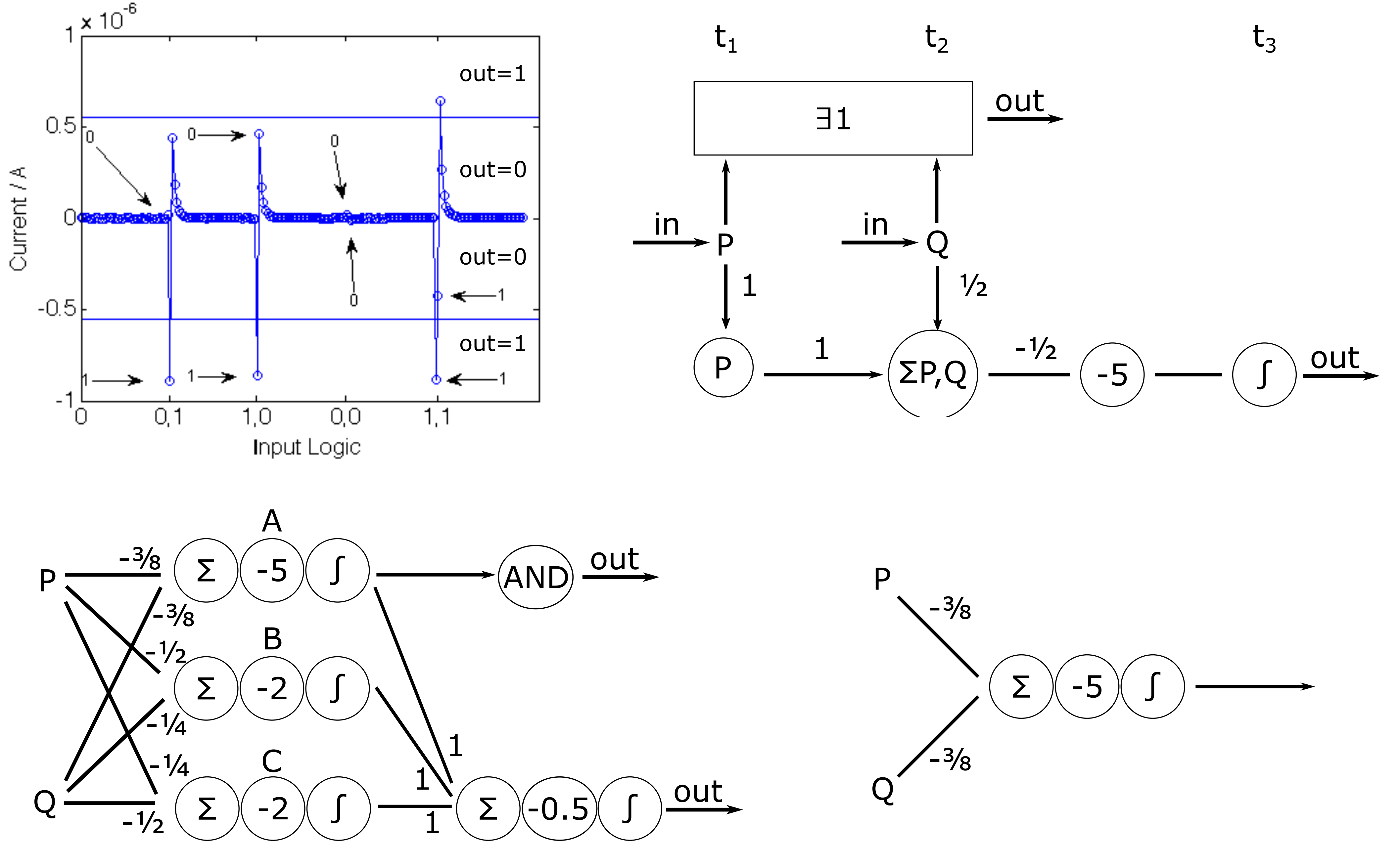}
  
  \rule[-.5cm]{0cm}{0cm}}
  \label{fig:AND}
  \caption{AND spiking memristor logic gate (SpMLG) and equivalent perceptron models. Top left (a): output from a SpMLG. The $\0$ (logical zero) and $\1$ (logical 1) are input as a small positive voltage spike and a large negative voltage spike, respectively. Outputs are the resulting current spikes, with a $\1$ taken as a current having a magnitude above $\pm 5.5\times 10^{-7}$A: the positive outputs give AND logic, the negative currents are inclusive OR. Top right (b): time-variant perceptron model for the process. $P$ is input at time-step 1 ($t_1$), $Q$ is input at time-step 2 ($t_2$), the output is read out in the positive current at time-step 3 ($t_3$). Both $P$ and $Q$ are taken as being the current measured $t_1$ and $t_2$ for ease of discussion. As the system `decays', the voltage input at $t_2$ gives only $\frac{1}{2}$ that at $t_1$, i.e. $Q= \frac{1}{2} P$, and the sum $P + Q$ decays by $\frac{1}{2}$ between $t_2$ and $t_3$, when it is read out. These decays give fractional weights on the diagram. If a $\1$ is input, a current which is less than -5.5$\times 10^{-7}$A is seen, which functions a $\1$-detector, $\exists \1$,  (which gives inclusive OR with this logic), and can happen at $t_1$ or $t_2$. Time is diagrammed as going from left to right, the computation proceeds from left to right along the `state' line, inputs and outputs are marked with small labelled arrows, events on the same vertical position happen at the same time, outputs resulting from an input appear above the computation line, those resulting from the memory appear below and the output of the computation appears on the far right. Bottom left (c): a standard (time-invariant) perceptron network that computes all the information computed by the SpMLG and, if $\1$ and $\0$ are -8$\times 10^{-7}$A and +1$\times 10^{-10}$A, the values entering the perceptron match the currents seen in the top left subfigure. Bottom right (d): a standard perceptron that can compute AND: this device predicts the incorrect current values for the perceptron inputs $a(\{\1 , \0\})$ and $a(\{ \0 , \1\})$ of $\sim+3\times 10^{-7}$A rather than $\sim+4\times 10^{-7}$A. The symbol $\sum$ indicates inputs, $x$, are summed, large circles with a number in indicate the bias value, $b$, applied, and $\int$ indicates that the perceptron will fire out a $\1$ if the value computed by $\sum x_i \cdot w_i + b > 0$ and will not fire if the value computed by $\sum x_i \cdot w_i +b \leq 0$.}
\end{figure}

An example of a SpMLG is given in figure \ref{fig:AND}a. The actual inputs are voltage spikes (as described in [13]), here we shall take the inputs, $x$, as the current associated with the voltage spike as found in a zeroed device, which is: $x(\1) = -8\times 10^{-7}A = -8u$ and $x(\0) =+1.2\times 10^{-10}A \approx 0u$ (and $u = 0.1\mu$A). We see from figure 2a that a second $\1$ input into the system has a different resulting current of $-4u$, not $-8u$, which we would expect if the inputs did not interact, or $-16u$ which we would expect if they were additive. This interaction is `diminishing returns', and it comes from the decay of a single spike. The output of $\{\0, \1\}$ is $\sim+4u$, and that of $\{\1,\1\}$ $\sim +6u$, which are recorded as a positive current response to setting the voltage to zero (`bounce-back'). A positive current recorded over $+5.5u$ is taken as a $\1$, that under $+5.5u$ is $\0$, in this way a threshold on the positive currents, or equivalently the current measured at $t_2$, would reproduce AND logic and this is similar to a standard perceptron AND gate. However, expanding the definition of $\1$ to include the negative currents over the threashold $\pm5.5u$ (and those $i<\pm5.5u$ are $\0$) demonstrates that the negative currents are performing inclusive OR logic, and thus the SpMLG can compute two logical functions on the same inputs and give the answers separately.

Figure~\ref{fig:AND}b is an attempt to describe the SpMLG's process with reference to the physical interactions. The SpMLG has only one input, so the two logical values are separated in time, but can interact due to the short-term memory of the memristor. Thus we separate out the 3 perceptron operations of summation, application of bias and thresholding (as described in the introduction) and allocate when they occur. A state line (bottom of Figure~\ref{fig:AND}b) indicates the computational steps as they occur in time, along with the state of the memristor. Inputs $P$ and $Q$ are drawn above the computation line and $P$ and $Q$ are strictly ordered in time so that $P$ is the input at $t_1$ and $Q$ is the input at $t_2$. The measured current of a time-step is an output, $y(t_i)$, and for the SpMLG, these acts as a $\1$-detector (this fires only when a $\1$ is detected), which computes inclusive OR. This detection can happen at either input step. This method of diagramming the process allows us to understand the observed current values. $Q(\1) = \frac{1}{2} P(\1)$ due to the interaction with the memory of the memristor, thus, $P(\1)=-8u$, $Q(\1)=-4u$. Summing these currents gives $-12u$ and we find that if the voltage is turned off the response spike that would be expected, if the system were linear, would be $+12u$. Instead, the value is $+6u$, so a weight of $\frac{1}{2}$ is applied to $P+Q$, accounting for the fact that an extra time-step is required which causes a second decrease in the values. Note that, in figure 2b, the threashold and $\int$ function is drawn on, although these functions are applied by inspection of the graph in figure 2a, not by the device itself. 

Having now gotten the `weights' for how the current values are combined in this SpMLG, we could try and draw a standard perceptron, but we have a problem. A standard perceptron that can perform a logical AND is given in figure 2d, for this system the input $\{\1, \1\} \rightarrow \1$ as $\frac{-3}{8} \cdot -8u = +6$, which is more than the bias, and so would correctly fire.  But if we look at the $\{\0,\1\}$ and $\{\1, \0 \}$ parts of the truth table (see table~\ref{tab:ANDMaths}), the $a$ values are incorrect as $\frac{-3}{8} \cdot -4u + \frac{-3}{8} \cdot 0u = +\frac{3}{2}$, which is less than the bias, but not equal to $+4u$ as in the actual system. To build a standard perceptron that is capable of computing AND the way the SpMLG does we need 3 perceptrons, as shown in figure 2c. Here, as $P$ and $Q$ are order-indepedent, to apply the time-dependent weighting on the second input, we need to compute the two possible ordered sums, and the correct sum has the correct a value. For example, if $PQ=\{\1,\0\}$, the sums computed at perceptrons A, B and C are: $\sum_A = 3u$, $\sum_B = 4u$ and $\sum_C = 0$, as the bias is $-3$, only perceptron $B$ would fire, and the $a$ values was correct. This perceptron network can also calculated the inclusive OR operation. This examination of the SpMLG demonstrates that it is equivalent to 3 standard perceptrons and more complicated than a normal perceptron, and gate as it cannot be precisely simulated by a single perceptron as an AND gate can. In fact, the single perceptron AND gate is logically equivalent to the SpMLG, as it will correctly perform AND logic, but not numerically equivalent, in that a single perceptron cannot account for the ordering of $P$ and $Q$ that is necessary. 

We can now associate the $a$ values (from equation~\ref{eq:a}) of a perceptron with a current,$i$, measured at a certain time (time-step j, $t_j$). Specifically, in a standard perceptron $a=\sum w_i \cdot x_i$, and in the memristor logic gate $a_j = i(t_j)$, so that the nodes in the perceptron diagram are related to time via a direct 1:1 mapping between the $a$ value entering a perceptron and the current observed in the memristor. This assumes that there are no hidden processes and that we can `see' the memristor's state. 

\begin{table}[t]
  \caption{Spiking memristor logic gate truth table (left 6 columns) and a binary full adder truth table (right 5 columns) for comparison. The threshold for AND is $>+5.5$, for OR $<-5.5$. All currents are in units of $1\times 10^{-7}$A}
  \centering
  \begin{tabular}{cc|cc|cc|cc|cc}
    \toprule
    \multicolumn{6}{l}{Spiking Memristor Logic Gate (SpMLG)} &  \multicolumn{4}{|l}{AND and OR truth table}        \\
    \cmidrule{1-6}
    \multicolumn{2}{c}{Inputs} &  \multicolumn{2}{|c|}{$a$ values} & \multicolumn{2}{|c|}{Outputs}         &
    \multicolumn{2}{c|}{Inputs} &  \multicolumn{2}{c}{Outputs}           \\
    \cmidrule{1-2} 
    $P$ 			& $Q$ 	& $w_p P$ 			& $w_Q Q$		& AND 		& OR &
    $P$ 			& $Q$ 	& AND		& OR		\\
    \midrule
		-8. 	& -8. 	& -8. 	& -4. 		&+6 	& -8 & $\1$ 	& $\1$ 	&
		$\1$	& $\1$	 \\
 		-8. 	& 0. 	& -8. 	& 0 	& +4 & -8 & $\1$ 	& $\0$  	&
 		$\0$	& $\1$	 \\
 		0. 	& -8 	& 0. 	& -8. 	& +4. & -8	& $\0$ 	& $\1$ & $\0$	& $\1$ 		\\
 		0	& 0 &   0 & 0 		& 0	& 0 & $\0$ 	& $\0$ 	&$\0$ &$\0$ \\
    \bottomrule
  \end{tabular}
  	\label{tab:ANDMaths}
\end{table}

\section{Full adder} 

With spiking memristor logic, it is also possible to make what has been called an `arithmetical full adder'. This can add up three spikes, which via the threasholding applied to the output positive values gives the arithmetical sum output, $y_s$= \{0, 1, 2, 3\}, of the binary input values, see figure 3a. This is an odd concept, usually a full adder outputs a binary sum of  the inputs that can only count up to 2 (i.e. $\1$), with the number 3 represented as $\1\1$ which is `3' in binary or a sum-bit and a carry-bit, as it is described in electronics. The negative currents gave extra information, spikes in region 2 of figure 3a were indicative of a carry bit (i.e. 2 or more $\1$'s input) and a current in region 1 of the graph were indicative of at least $\1$ having been input. 

\begin{figure}[ht!]
  \centering
  \fbox{\rule[-.5cm]{0cm}{0cm} 
  \includegraphics[width=0.9\linewidth]{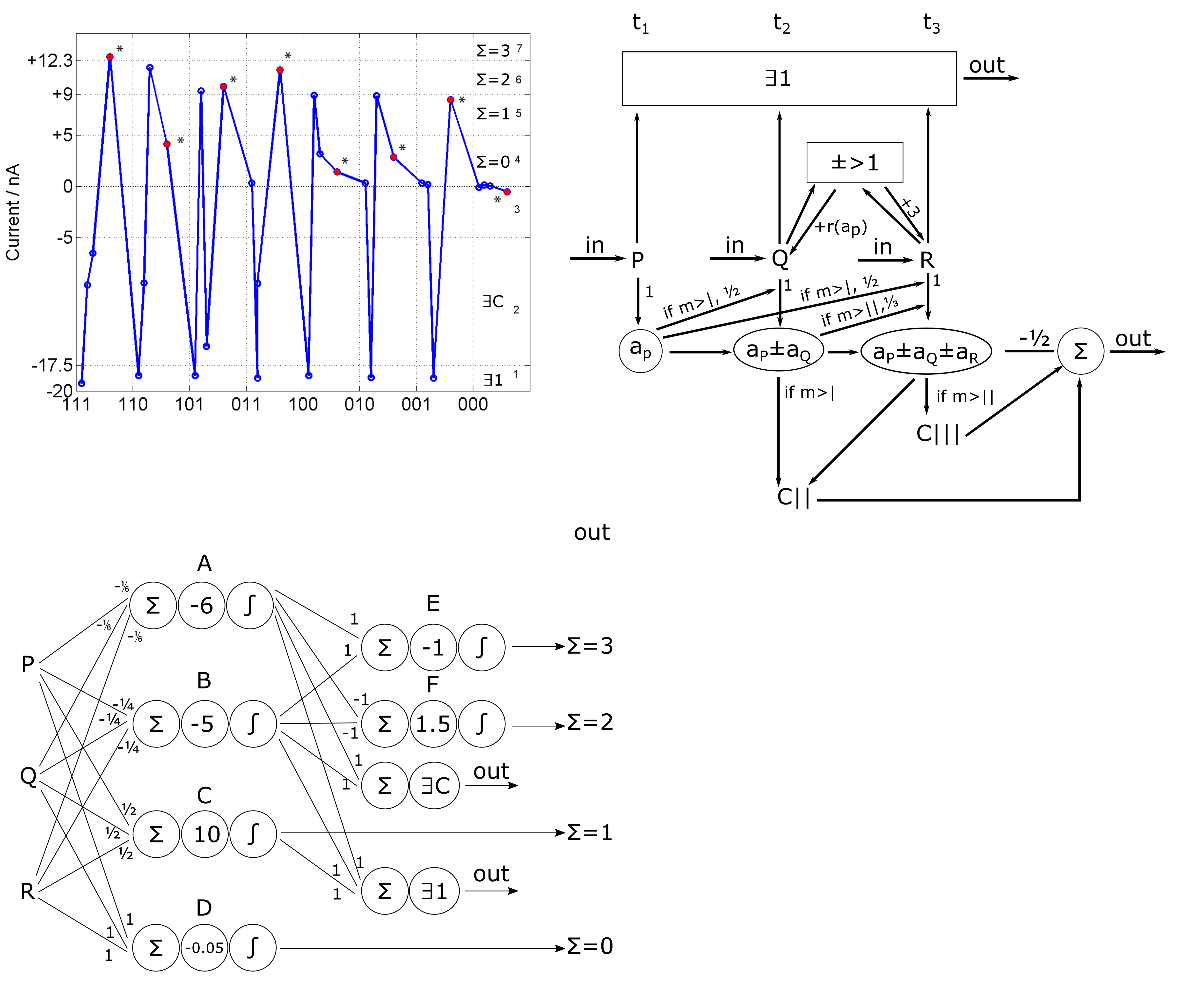}
  \rule[-.5cm]{0cm}{0cm}}
  \label{fig:FA}
  \caption{Arithmetic full adder (SpMLFA) made with a single memristor using spiking logic, and its perceptron equivalent models. Top left (a): outputs from a single memristor performing arithmetic full-adder addition. The arithmetic value of the sum, \{0, 1, 2, 3\} is encoded by the appearance of a spike current over the time period of the computation in areas \{4, 5, 6, 7\} of the plot, respectively. Area 1 functions as $\exists \1$, and a spike in area 2 indicates that the carry bit is $\1$. Note that $\1$ is taken as being the voltage that causes a current of -18nA in a fully zeroed device, $\0$ is the voltage that would cause a current of +0.05nA in a zeroed device. The memristor is zeroed in between each logical test to remove its short-term memory. Red points indicate the value at $t_3$. Top right (b): time-variant perceptron model for the SpMLFA. $R$ is input at time-step $t_3$. The $a$ values for $P$, $Q$ and $R$ are given by $a_P$, $a_Q$ and $a_R$ respectively. The values of the memristor's short-term memory, $m$, is marked on the computation line, and these can affect the weights of inputs if high energy values have been input, (i.e. any number of $\1$s). The sum is known on step $t_4$ from inspection of the maximum  positive current observed between $t_1$ and $t_4$. To get the actual current output value, two corrections must be applied dependent on whether the memory contains 2 or more $\1$s (which can be known on $t_2$ or $t_3$), i.e. if $m$ contains $\1\1$, a correction, c$\1\1\1$ must be applied, and similarly if $m=\1\1\1$ there is a second correction, c$\1\1\1$. The correction accounts for the energy lost from the memristor's state (its short-term memory) with each time-step, and relates to $m$ because this loss is only significant for states containing $\1$. The self-reflexive loop that accounts for the `bounce-back' effect seen when high energy state (one that contains a $\1$) exists and a $\0$ is input, followed by a $\1$ (i.e the system crosses 0V twice with high energy, as designated by $^{+}/_{-}>1$)--this only happens in this truth table at $\{\1 , \0 , \1 \}$. Outputs resulting from an input appear above the computation line, those resulting from the memory, $m$, appear below and the output of the computation appears on the far right. Bottom left (c): a standard (time-invariant) perceptron network equivalent to the arithmetic FA. The weights applied to the inputs will match the currents recorded in the top left figure. Note that it is possible to make a hybrid network where the outputs of the hidden layer neurons used as the current values by not applying a thresholding function.}
\end{figure}

Understanding and diagramming the operation of this sort of device is a little more involved that the SpMLG. For the numerics of this device, we now take $u$ to be 1nA, and $x(\1) = -18u$ and $x(\0)=+0.05u$, which we could approximate as zero as we did above, but here we keep the actual value. A rough measure of the magnitude of the sum is $\sum \{P, Q, R \} = x + \frac{x}{2} + \frac{x}{3}$. It seems that this takes into account the effect of a device having a short-term memory on the response to the input, and the weights, $\{\frac{1}{1},\frac{1}{2}, \frac{1}{3}\}$ suggests that they follow the function $w(t_n)=\frac{1}{n}$, which appears to match the curve in figure 1. Thus, if $P$, $Q$ and $R$ are strictly entered in that order then: $a_P = P$, $a_Q=\frac{1}{2}P$ and $a_R=\frac{1}{3}R$. The actual input values are also decreasing, and this involves adding the corrections of $c_{\1\1} = -\frac{1}{3}\frac{x(\1)}{2}$ (or $\frac{-a_Q}{3}$ and nominally associated with the decay of the second $\1$ input value at $t_3$) and $c_{\1 \1 \1} = -\frac{1}{3} \frac{1}{2}\frac{x(\1)}{3}$ (or $\sim -\frac{1}{3}\frac{1}{2} a_R$) nominally associated with the $P$ value being held for 2 input steps. As the $x(\0)$ is very small and energetically below that required to measurably change the memristor's state, we can ignore $\0$ inputs to the energetic corrections as was done for the AND gate, and is done in figure 3b. A more complete description including the $\0$ contributions is given in equation~\ref{eq:FA} and table\ref{tab:FAMaths}, which models the FA SpMLG to within our desired accuracy. 

\begin{equation}
\frac{a_P \pm a_Q \pm a_R}{2} + \frac{1}{3} Median[\{a_P, a_Q, a_R\}] + \frac{1}{6} Min [\{|a_P|, |a_Q|, |a_R| \}]
\label{eq:FA}
\end{equation}

Note that equation \ref{eq:FA} requires the use of Median and Min functions to sort the value by magnitude. As the effect of time and order is important to understanding the system, the SpMLFA is better understood as a Turing machine where values can be input and output from the memory and the value of the memory can affect the inputs to it. A very simple simulation following the design in figure~\ref{fig:fa}b was written to verify this design.

Table~\ref{tab:FAMaths} illustrates some interesting points about the SpMLFA. We are interested in understanding the behaviour, rather than precisely modelling this particular device, so we approximate the output currents seen in Figure 3a to the arithmetical sum, as the values $\sim+12.5u$ for 3, $\sim+10.5u$ for 2, $\sim+9$ for 1 and $\sim0$, these came from approximating the $a_P$, $a_Q$ and $a_R$ as integers: \{18, 9 and 6\}. If $\0$ is approximated as $0u$, then we get these approximations from the equation. The inclusion of the actual value for the $\0$ moves those values slightly, but is both close enough (the output currents are within the ranges) and demonstrates why there is a slight difference in output sum between arithmetically equivalent lines of the truth table (e.g. between the outputs of $\{\0,\1,\1\}$, $\{\1,\0,\1\}$ and $\{\1,\1,\0\}$). In order to get the correct $a$ values, we see the unexpected  value of $+20u$ in line 3 of table~\ref{tab:FAMaths}. The rules for calculating the $a$-values outlined above do not take into account the effect of `double-bounce-back' which is seen when the memory contains a $\1$ (i.e. has sufficient energy) goes to a $\0$ and then a $\1$ before being zeroed--this changes the sign in the sum and this is indicated in figure~\ref{fig:FA}b as a self-reflexive loop and the $\pm$ in the sum. The transition from $\1 \rightarrow \0$ yields a $+9u$, the transition from $\0 \rightarrow \1$ yields a $-15$ before giving an output summation of $\sim10u$. When the system crosses 0 the contents of the memory is output as a response spike. So when we go from $\1 \rightarrow \0$ the response spike includes the value of $m$, which is $-9u$, so $a_Q$ is really $+0.05 + 9u$. There is `friction' associated with switching sign, this is why $R$ for $\{\1, \0, \1\}$ is $\sim -15u$ rather than $-18u$. This effect is rendered in the diagram in figure 3b as the self-reflexive loops and is only significant for $\{\1, \0, \1\}$.

\begin{table}[t]
  \caption{Spiking memristor logic arithmetical full-adder table (left 6 columns) and a binary full adder truth table (right 5 columns) for comparison.}
  \centering
  \begin{tabular}{lll|lll|lcc|ccc|cc}
    \toprule
    \multicolumn{9}{l}{Arithmetical full adder} &  \multicolumn{5}{|l}{Binary full adder}        \\
    \cmidrule{1-9}
    \multicolumn{3}{c}{Inputs} &  \multicolumn{6}{c}{Outputs}         &
    \multicolumn{3}{c}{Inputs} &  \multicolumn{2}{c}{Outputs}           \\
    \cmidrule{1-3} 
    	&  &  & \multicolumn{3}{c}{$a$ values} & sum & carry & &
    	&  &  & sum & carry  \\
    $P$ 			& $Q$ 	& $R$ 	& $w_p P$ 			& $w_Q Q$ 	& $w_R R$ 	& $\sum$		& $C$ 		& $\exists \1$ &
    $P$ 			& $Q$ 	& $R$ 	& $\sum$		& $C$ 		\\
    \midrule
		-18. 	& -18. 	& -18. 		& -18. 	& -9. 	& -6. 		&+12.5 	& $\1$ 	& $\1$ 	&
		$\1$	& $\1$	& $\1$ 	& $\1$ 	& $\1$ 	 \\
 		-18. 	& -18. 	& +0.05 	& -18. 	& -9. 	& +0.05 	& +10.5	& $\1$ 	& $\1$  	&
 		$\1$	& $\1$	& $\0$ 	& $\0$ 	& $\1$ 	 \\
 		-18. 	& +0.05. 	& -18. 	& -18. 	& +9.05. 	& -15. 	& +10.47 	& $\1$ 	& $\1$ 	&
 		$\1$	& $\0$	& $\1$ 	& $\0$ 	& $\1$ 	\\
 		+0.05 	& -18. & -18. 		&+0.05 	& -18. & -9. 		& +10.7	& $\1$ 	& $\1$  	&
 		$\0$	& $\1$	& $\1$ 	& $\0$ 	& $\1$ 	\\
 		-18. 	& +0.05 & +0.05 	& -18. 	& +0.05 & +0.05 	& +8.9 	& $\0$ 	& $\1$ 	&
 		$\1$	& $\0$	& $\0$ 	& $\1$ 	& $\0$ 	\\
 		+0.05 	& -18. & +0.05 	& +0.05 	& -18. & +0.05 	& +8.9 	& $\0$ 	& $\1$ 	&
 		$\0$	& $\1$	& $\0$ 	& $\1$ 	& $\0$ 	\\
 		+0.05 	& +0.05 & -18. 	& +0.05 	& +0.05 & -18. 	& +8.9 	& $\0$ 	& $\1$ 	&
 		$\0$	& $\0$	& $\1$ 	& $\1$ 	& $\0$ 	\\
 		+0.05 	& +0.05 & +0.05 	& +0.05 	& +0.05 & +0.05 	&-0.1 		& $\0$ 	& $\0$ 	&
 		$\0$	& $\0$	& $\0$ 	& $\0$ 	& $\0$ 	\\

    \bottomrule
  \end{tabular}
  	\label{tab:FAMaths}
\end{table}

The standard perceptron that would be capable of reproducing this system with the same $\{a_i\}$ is given in figure 3b, and requires 6 hidden layer neurons. Interestingly, if the threasholding was removed from the output channels, this network could give the $y_s$ as the current values observed in the SpMFA, as continuous values, separately from the binary outputs of the carry bit and $\exists \1$ `gate'. Figure 2d shows a standard perceptron binary 2-bit full adder. As above, a comparison between 3c and d demonstrates that the SpMLFA is more complex than a binary full adder.   

\section{Discussion}

In this paper, we have seen that a novel device called the memristor natively spikes in response to a change in voltage, and that these spikes can interact in an interesting, non-linear manner that allows the computation of binary logical operations and arithmetical sums. These memristors operating as gates do `through-time' computation using the memristor's short-term memory (or state) to hold previously input values. This is a different mode of operation to standard perceptrons which primarily compute `through space' via the connections between different perceptrons. In comparing the two, we see that the space-based $a$ values which are the sums of inputs and input weights are directly equivalent to the time-based $a$ values measured from the spiking memristor as data in input: and thus that a memristor is a type of perceptron. However, a standard, simple perceptron has no way of enforcing the order of inputs, and should be order-invariant to the inputs (the order of the sum does not matter). Order-Invariance is time-invariance if the order is defined over time. The memristors is not order-invariant because inputs arriving at different times find the memristor in a different state and thus get a different response. As the memristor state decays non-linearly, the ordering in enforced and observable. 


An interesting point about perceptrons is that although they may take binary inputs and produce binary outputs, they do this with access to the entire real number space: any value of weights can be used and thus the a values can be any real number. With application of a threasholding rule, these values are `projected down' to binary number space. The memristor logic gates presented here could do the same thing, if threasholds were applied to the numerical output (the measured currents), then they would operate like a binary perceptron. Not applying this threasholding allows the computation of actual values, which, with the correct choice of weights, could allow analogue computation across a network.  


I made the point that (biological) neurons have been described as living memristors, can this work suggest anything about living neural networks? Real neurons have a refractivity period, which the memristors also has--it requires some time to return to a zeroed state. Memristors also require inputs to arrive within a certain time window (whilst previous inputs are stored in the short-term memory) and, these interactions fall off rapidly as in spike-time-dependent plasticity. It may well be that living neurons are best described as time-variant perceptrons, rather than time-invariant perceptrons, and if so, then memristors would be a good choice of artificial spiking neurons for neuromorphic computing.  Also, as the memristors are low power consumption and operate with physiological currents, they might even been good components for connecting to biological neural networks. In our further work, we shall investigate whether some of these ideas apply to models of neural networks, and whether time-varying perceptron models can help explain perception-related errors.





\subsubsection*{Acknowledgments}

Authors acknowledge Levehulme Trust grant number RPG-2016-113.

\section*{References}

\small

[1] D. B. Strukov, G. S. Snider, D. R. Stewart\ \& R. S. Williams (2008) The missing memristor found. {\it Nature}, {\bf 453}, pp. 80--83.
  
[2] S. H. Jo, T. Chang, I. Ebong, B. B. Bhadviya, P. Mazumder\ \& W. Lu.\ (2010)
Nanoscale memristor device as a synapse in neuromorphic systems. {\it Nanoletters}, {\bf 10}, pp. 1297--1301.

[3] C. Zamarreno-Ramos, L. A. Carmu\~{n}as, J. A. P\'{e}rez-Carrasco, T. Masquelier, T. Serrano-Gotarredona\ \& Bernab\'{e} Linares-Barranco,\ (2011) On Spike-timing dependent plasticity, memristive devices and building a self-learning visual cortex, {\it Frontiers in Neuormorphic engineering},\ {\bf 5},\ pp. 26(1)--26(20).

[4] G. D. Howard, E. Gale, L. Bull, B. de Lacy Costello\ \& A. Adamatzky.(2012) Evolution of Plastic Learning in Spiking Networks via Memristive Connections. {\it IEEE Transactions on Evolutionary Computation}, {\bf 16}, pp. 711--719.

[5] G. D. Howard, L. Bull, B. de Lacy Costello, E. Gale\ \& A. Adamatzky. (2014) Evolving spiking networks with variable resistive memories, {\it Evolutionary computation}, {\bf 22}, pp. 79--103.

[6] L. O. Chua\ \& S. M. Kang. (1976) Memristive devices and systems. {\it Proceedings of the IEEE}, {\bf 64}, pp. 209--223.
  
[7] L. Chua. (2013) Memristor, Hodgkin–Huxley, and Edge of Chaos. {\it Nanotechnology}, {\bf 24}(38), 383001.
  
[8] E. Gale. (2014) Memristors and ReRAM: Materials, Mechanisms and Models (a review). {\it Semiconductor Science and Technology}, {\bf 29}, 104004.

[9] E. Gale, B. de Lacy Costello\ \& A. Adamatzky. (2013), Observation, Characterization and Modeling of Memristor Current Spikes.  {\it Applied Mathematics and Information Sciences}, {\bf 7}, 1395--1403.

[10] E. Gale, B. de Lacy Costello, V. Erokhin\ \& A. Adamatzky. (2014) The Short-term Memory (d.c. response) of the Memristor Demonstrates the Causes of the Memristor Frequency Effect. In Proceedings of CASFEST 2014, IEEE Press

     
[11] D. Gater, A. Iqbal, J. Davey\ \& E. Gale. (2013) Connecting Spiking Neurons to a Spiking Memristor Network Changes the Memristor Dynamics. In {2013 International Conference on Electronics, Circuits and Systems (ICECS)}, pp. 534--537, IEEE Press.

[12] E. Gale, D. Pearson, S. Kitson, A. Adamatzky\ \& B. de Lacy Costello. (2015) The effect of changing electrode metal on solution-processed flexible titanium dioxide memristors, {\it Materials Chemistry and Physics}, {\bf 162}, 20--30.

[13] E. Gale, B. de Lacy Costello\ \& A. Adamatzky. (2013) Boolean Logic Gates from a Single Memristor via Low-Level Sequential Logic. In G. Mauri, A. Dennunzio, L. Manzoni\ \& A. E. Porreca (eds.), {\it Unconventional Computation and Natural Computation 7956}, Lecture Notes in Computer Science (series), pp.\ 79--89. Springer Berlin Heidelberg.
  
[14] E. Gale, B. de Lacy Costello\ \& A. Admatzky. (2013) Is Spiking Logic the Route to Memristor-Based Computers? In {\it 2013 International Conference on Electronics, Circuits and Systems (ICECS)}, pp. 297--300. IEEE Press.

\end{document}